\documentclass[prb,twocolumn,showpacs,showkeys,amsmath,amssymb]{revtex4}

\usepackage{graphicx}
\usepackage{dcolumn}
\usepackage{bm}
\usepackage{courier}

\begin{document}


\title{A Note on the Calculation of Averages in Superconducting Cable-in-Conduit Conductors}


\author{A.Anghel}
\email[]{alexander.anghel@psi.ch}
\affiliation{Paul Scherrer Institut, CH-5232 Villigen PSI}


\date{\today}

\begin{abstract}
We show that there are two different ways of calculating the
average electric field of a superconducting cable in conduit
conductor depending on the relation between the current transfer
length and the characteristic self-field length.
\end{abstract}


\maketitle

\section{Introduction}

Measuring the volt-ampere characteristic of
superconducting cable-in-conduit conductors is one of the most
important way  to completely characterize their DC behavior. In
order to measure the voltage drop along the conductor, voltage
taps are placed on the conduit (a Ti or stainless-steel jacket) at
some distance apart in the high field region of the sample. This
distance should be greater than the length of the twist pitch of the last
cabling stage in order to comply with the request of complete transposition of strands.

\begin{figure}[tb]
\centerline{\includegraphics[width=3in,keepaspectratio]{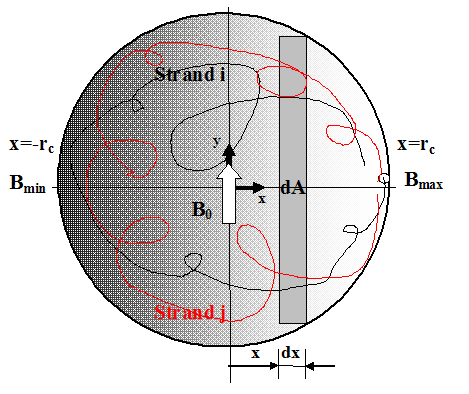}}
\caption{\label{fig1} Two typical strand trajectories in a given section of the cable
with a field gradient as a result of the superposition of the uniform background field with the self-field of the cable. In this configuration the field gradient points in the Ox direction.}
\end{figure}

The superposition of the magnetic field produced by
the transport current in the sample and the external background
field $B_0$ results in a linear magnetic field gradient in the sample
cross-section as shown in Fig.\ref{fig1}. 

On a certain  line the magnetic field will vary between a minimum $B_{min}$ and a maximum value $B_{max}$. Each strand among the thousands of strands in the
cable 'travels' on a complicated spiraled path trough
this field gradient and has therefore different locally defined critical
currents. Two typical strand trajectories are shown in Fig.\ref{fig1}. Assuming that all strands are charged uniformly with
current by a perfect joint, as soon as the strands arrive in the high field region of the cable,
the condition is met that some strands will happen to carry a current higher that the local
critical current imposed by the local magnetic field in the cable at this position. Depending on the degree of
transversal conductance between the  strands, a more or less intensive
current transfer process starts. If the transversal conductance is high and the length of the strand path is large, at steady state the current in each strand will be modulated following more or less the magnetic
field pattern seen by the strand on its path. The measured
electric field is then an average of the electric fields generated by
the strands in the cable.  The average electric field can be calculated by integrating the electrical field along the length of the strands. Assuming that the cable is ergodic \cite{ergodic} 
this average is equal with the cable cross-section ensemble average calculated using the strand
geometrical probability distribution \cite{pdf}. This
geometric averaging and some subtleties involved in their calculation is the object of the present investigation.

\section{Two characteristic lengths in cable}
Excepting the cables made of insulated strands, in real cables there is always a certain current transfer possible. Depending on the inter-strand transversal resistance, the current transfer length $L_{CT}$ is defined as the length needed to
balance an initial current inhomogeneity. Extending (without
proof) to a full-size cable the relation proposed earlier by Ekin
\cite{Ekin} for the filaments in a strand, we can define a current-transfer length $L_{CT}$ by 

\begin{equation}
\label{eq2}
L_{CT}=\left(\dfrac{0.1}{n}\right)^{\frac{1}{2}}\left(\frac{\rho_t}{\rho_c}\right)^{\frac{1}{2}}D_c
\end{equation}

where $D_c$ is the cable diameter, $\rho_t$ is the transversal
resistivity, a measure of the inter-strand contact
resistance and $\rho_c$ is the resistivity criterion, $\rho_c=E_cJ_c$ defined with the help of the electric field criterion $E_c$ and the critical current density $J_c$. $n$ is the power-law index from the nonlinear current-voltage characteristic of the superconducting strands

\begin{equation}
\label{powlaw}
E=E_c\left(\dfrac{J}{J_c}\right)^n
\end{equation}

The strand position in the cable cross-section $P(x(z),y(z))$ can be described with a very good accuracy by the following equations:

\begin{eqnarray}
\label{traj}
x_i(z)=\sum_{k=1}^S r_k \cos(\dfrac{2\pi z}{p_k}+\phi_i) \nonumber \\
y_i(z)=\sum_{k=1}^S r_k \sin(\dfrac{2\pi z}{p_k}+\phi_i)
\end{eqnarray}

where: $z$ is the axial coordinate along the cable axis, $r_k$ and $p_k$ are the radius and the  twist-pitch of the $k$-th stage and $S$ the number of stages. The index $i\in 1..N$ and the phases $\phi_i$ are introduce to describe different strands with different initial positions. $N$ is the number of strands in the cable. We adopt here the convention that the twist-pitch increases with the $k$ index, the smallest being $p_1$ and the greatest $p_{last}=p_S$. If we follow and record the position of one strand at enough many slices at coordinate $z_n=np_S$, an integer multiple of the last twist-pitch length, we will see that these recorded positions cover almost uniformly the cross-section of the cable. Moreover, if we look at all strands position in one slice we will see a similar uniform distribution. From a statistical point of view we can not distinguish between the two pictures. In this case we say that the cable is ergodic and we can replace the average over the length of one strand by an average over all strands in one slice. Stated in a more simple way, the average over the length $L$ of any physical cable property $X$ can be replaced by a cross-section average with a suitable probability distribution function $w(x,y)$

\begin{equation}
\label{ergodic}
\left\langle X \right\rangle=\dfrac{1}{L}\int X(s)ds=\int w(x,y)X(x,y)dxdy
\end{equation} 

with $s$ the coordinate along the cable length and $\left\langle X \right\rangle$ is the average of $X$.

The fact that the cable cross-section is circular and is covered by a magnetic field distribution in the form of a linear gradient as shown in Fig.\ref{fig1} forces us to calculate the probability distribution function of the strands  having coordinate between $x$ and $x+dx$ but an arbitrary value of $y$. All this strands feel the same magnetic field. This new distribution function can be calculated with the relation

\begin{equation}
\label{prob}
w(x)dx=\dfrac{dN(x)}{N}
\end{equation}

where $dN(x)$ is the number of strands having coordinate between $x$ and $x+dx$ and $N$ the total number of strands in the cable. If the strands are uniformly distributed in the cable cross-section with a density $n_0$, $dN(x)$ and $N$ are proportional to the area of the stripe of width $dx$ (Fig.\ref{fig1}), $dN(x)=n_0dA=2n_0\sqrt{r_c^2-x^2}dx$ and the total cross-section area of the cable, $N=n_0A=n_0\pi r_c^2$.

After substitution in Eq.\ref{prob} one obtains

\begin{equation}
\label{pxfinal}
w(x)=\dfrac{2\sqrt{r_c^2-x^2}}{\pi r_c^2}
\end{equation}

\begin{figure}[tb]
	\centering
		\includegraphics[width=3.5in,keepaspectratio]{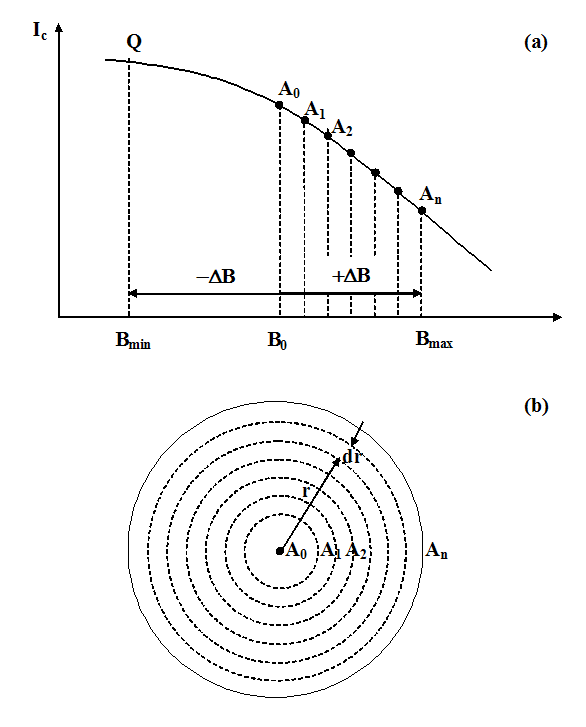}
		\caption{\label{fig2} Effect of magnetic field gradient in the cable cross-section on the critical current of strands. }
\end{figure}

If the twist pitch length of the last stage $p_{last}$ is long, the strands on the high field region of the cross-section field gradient will have a local critical current which is lower than the carried current and the strands on the low field region of the gradient, a critical current higher than the own current. This is illustrated in Fig.\ref{fig2}a.

In this case the averages can be calculated with a different probability function, p(B) which can  be calculated with the help of the probability distribution $w(x)$. 

Let us represent the magnetic field distribution in the cable as

\begin{equation}
\label{field}
B(x)=B_0+kx
\end{equation}

where $B_0$ is the background magnetic field and $k$ a proportionality constant. As shown in Fig.\ref{fig1}, $B_{min}=B(-r_c)=B_0-kr_c=B_0-\Delta B$ and $B_{max}=B(r_c)=B_0+kr_c=B_0+\Delta B$. We treat now $x$ as a random variable defined on the interval $[-r_c,r_c]$ with the distribution $w(x)$ given by Eq.\ref{pxfinal}. By virtue of Eq.\ref{field}, B(x) is also a random variable. Let us calculate now the probability distribution function of $B(x)$, a function of the random variable $x$ with probability distribution function $w(x)$. We start with the standard definition of a probability distribution of a random variable which is a function of another random variable with known distribution 

\begin{eqnarray}
\label{newdistr}
p(B)=\int_{-r_c}^{r_c} w(x) \delta (B-B(x))dx=\nonumber \\
=\int_{-r_c}^{r_c}\dfrac{2\sqrt{r_c^2-x^2}}{\pi r_c^2}\delta (B-B(x))dx
\end{eqnarray}

The integral in Eq.\ref{newdistr} can be calculated with the help of the following known relation in the theory of $\delta$-distribution function

\begin{equation}
\label{delta}
\int f(x)\delta(g(x))dx=\sum_n \dfrac{f(x_n)}{\left| \dfrac{\partial g}{\partial x} \right|_{x=x_n}}
\end{equation}

where $x_n$ are the roots of the equation $g(x)=0$. In our case, $g(x)=B-B(x)=B-B_0-kx=0$, has a single root $x_1=(B-B_0)/k$ and $\partial g/\partial x=k$. Substituting this in Eq.\ref{newdistr} we obtain

\begin{eqnarray}
\label{newdistr1}
p(B)=\dfrac{2\sqrt{r_c^2-x_1^2}}{\pi r_c^2}\dfrac{1}{k}=\nonumber \\
=\dfrac{2}{\pi kr_c^2}\sqrt{r_c^2-\left(\dfrac{B-B_0}{k}\right)}=\nonumber \\
=\dfrac{2}{\pi (\Delta B)^2}\sqrt{(\delta B)^2-(B-B_0)^2}
\end{eqnarray}

This function is the probability distribution of the field $B\in [B_{min}, B_{max}]$. If we perform a change of variable $B-B_0\rightarrow B $, the probability distribution function can be written as

\begin{equation}
\label{pbfinal}
p(B)=\dfrac{2}{\pi (\Delta B)^2} \sqrt{(\Delta B)^2-B^2}
\end{equation}

where now the variable $B\in[-\Delta B, \Delta B]$.

Let us apply this relation to the calculation of the average critical current in a cable. From Eqs.\ref{ergodic} and \ref{pxfinal} we have  

\begin{eqnarray}
\label{avgic}
\left\langle I_c \right\rangle=\int_{-\Delta B}^{\Delta B} p(B)I_c(B_0+B)dB=\nonumber \\
=\int_{-\Delta B}^{\Delta B} \dfrac{2\sqrt{(\Delta B)^2-B^2}}{\pi (\Delta B)^2}I_c(B_0+B)dB
\end{eqnarray}

If the twist pitch is shorter than the current-transfer length, the current-carrying capacity of the strand will be determined by the critical current at the most highly field in the cable cross-section. The strand who crosses the $B_{max}$ position at $x=r_c$ will keep its current $I=I_c(B_{max})$ all over the way down from that section and will create a circular region of constant current $A_n$. The same is true for all other strands who pass a region with a field intensity $B_0<B<B_{max}$ thus creating circular regions of constant current $A_{n-1}, A_{n-2},...,A_2, A_1, A_0$ as shown in Fig.\ref{fig2}b. The linear field gradient on the cable cross-section, created by the overlapping of the self and background fields, is replaced by a circular current distribution. Applying Eq.\ref{prob} to this case and observing that $dN(x)=n_02\pi xdx$ and $N=n_0\pi r_c^2$ we obtain the distribution function for this case

\begin{equation}
\label{pcirc}
w(x)=\dfrac{2x}{r_c^2}
\end{equation}
 
where $x\in[0,r_c]$. The probability distribution of the field can be calculated similar with the calculation performed before for the long twist pitch case. We have

\begin{eqnarray}
\label{pb2}
p(B)=\int_0^{r_c}w(x)\delta(B-B(x))=\nonumber \\
=\dfrac{2(B-B_0)}{(kr_c)^2}
\end{eqnarray}

with $B$ in the range $[B_0, B_{max}]$. If we change the field variable $B-B_0\rightarrow B$ and observe that $kr_c=\Delta B$ we can write the probability distribution as

\begin{equation}
\label{pb2new}
p(B)=\dfrac{2B}{(\Delta B)^2}
\end{equation}

where $B\in[0,\Delta B]$. The average of critical current calculated with this distribution is

\begin{equation}
\left\langle I_c \right\rangle=\int_0^{\Delta B} \dfrac{2B}{(\Delta B)^2} I_c(B_0+B)dB
\end{equation}

and applies if the last twist pitch of the cable is shorter than the current-transfer length. 

\section{Including the angular distribution in the average process}
\begin{figure}[tb]
\centerline{\includegraphics[width=3in,keepaspectratio]{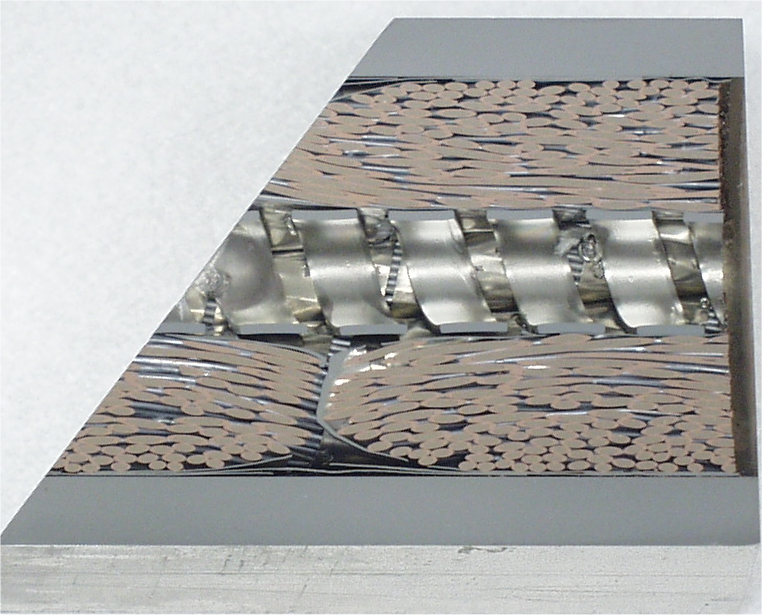}}
\caption{\label{fig3} A longitudinal cut in a cable in conduit conductor. The cross-section of the cut in the strands is decreasing radially from center to the edge.}
\end{figure}

The average calculated with the geometrical probability
distribution function does not take into account the fact that not
all strands cut the cable section at the same angle. A
longitudinal cut trough a cable reveals this problem. As shown in
Fig.\ref{fig3}, the cut-off cross-section of strands close to the cable
center is very large (elongated ellipses) indicating that these strands are almost
parallel with the cable axis. Moving away from cable axis in the radial direction, the cross-section area of the strand-cuts decrease reaching a minimum (almost circular) at the cable edge, where the strands are almost perpendicular to the cable axis. 

Therefore, besides the uniform distribution of strand position in
the cable cross-section which leads to the geometrical
probability distribution $w(x)$ of Eq.\ref{pxfinal}, we have also a radial distribution of
strand angles relative to the cable axis. Both distribution overlap, coexist and are interrelated. The cable phase-space is therefore not restricted to the set of points $\left\{x_i(z), y_i(z)\right\}_{i\in \textbf{N}}$. It should be extended to the set $\left\{x_i(z), y_i(z), x'_i(z), y'_i(z)\right\}_{i\in \textbf{N}}$ with $x'_i=dx_i(z)/dz$ and $y'_i(z)=dy_i(z)/dz$, where $x'_i$ and $y'_i$ are the tangent of the angles the strand makes with the $Ox$ and $Oy$ axis.

\begin{eqnarray}
\label{tangents}
\tan\theta_{x,i}(z)=x'_i(z) \nonumber \\
\tan\theta_{y,i}(z)=y'_i(z)
\end{eqnarray}

In this section a new average formula will be developed which accounts for this extension.

\begin{figure}[tb]
\centerline{\includegraphics[width=3.4in,keepaspectratio]{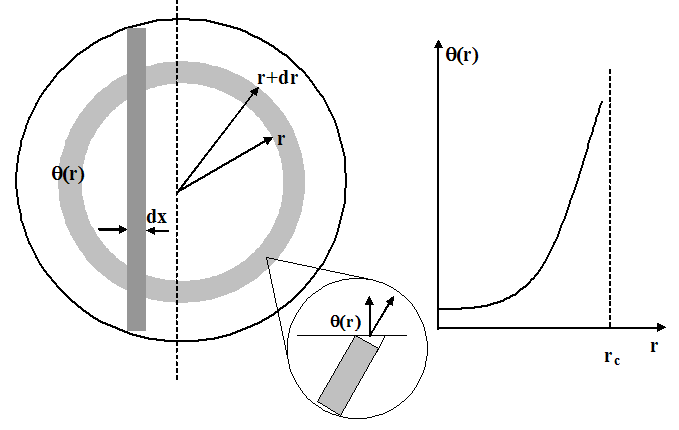}}
\caption{\label{fig4} The distribution of strand cut-angles in a typical cable in conduit conductor.}
\end{figure}

Usually, one calculates the average electric field by
considering a geometry as in Fig.\ref{fig1} where the
vertical stripe of width $dx$ includes all strands sensing the
same local magnetic field $B(x)$. The strand cut-angle has a radial distribution as illustrated in Fig.\ref{fig3}. The angular average is calculated using a set of concentric rings of width $dr$ containing strands having the same cut-angle $\theta (r)$ as shown in Fig.\ref{fig4}. The variation of the cut-angle with the radius is encoded in the $\theta$ dependence of $r$, a smooth convex function. In order to calculate the probability distribution function, we make a change of variable from initial
variables $x,y$ to the variables $x,r$. The transformation is

\begin{equation}
\label{eq4} 
x\equiv x(r,x)=x; \quad \quad  y\equiv
y(r,x)=\sqrt{r^2-x^2}
\end{equation}

The Jacobian of this coordinate transformation is

\begin{eqnarray}
\label{eq5} 
\mathcal{J}=\left|%
\begin{array}{cc}
  \dfrac{\partial x}{\partial x} & \dfrac{\partial x}{\partial r} \\ \\
  \dfrac{\partial y}{\partial x} & \dfrac{\partial y}{\partial r} \\
\end{array}%
\right| =\left|\begin{array}{cc}
                           1 & 0 \\ \\
  -\dfrac{x}{\sqrt{r^2-x^2}} & \dfrac{r}{\sqrt{r^2-x^2}} \\
\end{array}\right|= \nonumber \\
=\dfrac{r}{\sqrt{r^2-x^2}} \nonumber \\
\end{eqnarray}

The average electric field $\left\langle  E\right\rangle$ is calculated with the basic relation

\begin{equation}
\label{eq6} 
\left\langle E\right\rangle=\int \dfrac{dA}{A}E(x,y)=\int \dfrac{dxdr}{A}E(x,r,\theta)\mathcal{J}
\end{equation}

where $A$ is the cross-section area of the cable. If we use the Jacobian, Eq.\ref{eq5} we get

\begin{eqnarray}
\label{eq7}
\left\langle  E\right\rangle= \dfrac{1}{A}\int dx dy E(x,y,\theta)= \nonumber \\
=\dfrac{2E_c}{\pi r_c^2}\int_{-r_c}^{r_c} dx \int_x ^{r_c}
\dfrac{rdr}{\sqrt{r^2-x^2}}\left( \dfrac{J}{J_c(x)}
\cos(\theta(r))\right)^n \nonumber \\
\end{eqnarray}
where we have taken into account that only the normal component of the current density $J\cos(\theta)$ contributes to the axial electric field.
Separating the variables one finally obtains

\begin{equation}
\label{eq8} 
\left\langle E\right\rangle=\dfrac{2E_c}{\pi r_c^2} \int_{-r_c}^{r_c} dx
\left[\dfrac{J}{J_c(x)}\right]^n \int_x^{r_c}dr
\dfrac{r\cos^n(\theta (r))}{\sqrt{r^2-x^2}}
\end{equation}

This formula has a very simple structure. The average electric
field is first the "sum" of local electric fields in the cable
cross-section of constant field times a weighting factor $w(x)$
which is contained in the second (the $r$-integral). The limiting case is when all strands are parallel to the cable axis. Then, $\theta(r)=0$ for all $r\in [0,r_c]$ and  we get

\begin{equation}
\label{eq9} 
w(x)=\dfrac{2}{\pi r_c^2}\int_x^{r_c}dr
\dfrac{r}{\sqrt{r^2-x^2}} = \dfrac{2\sqrt{r_c^2-x^2}}{\pi r_c^2}
\end{equation}

which is the well known geometric probability distribution
function \cite{pdf} for a simply connected cable with circular cross-section.

In other words we can keep the standard formula for the average
electric field

\begin{equation}
\label{eq10} 
\left\langle E\right\rangle=\int_{-r_c}^{r_c}w(x)E(x)dx
\end{equation}

but with a modified weight or probability distribution function

\begin{equation}
\label{eq11} 
w(x)=w_{\theta}(x)=\dfrac{2}{\pi r_c^2}
\int_x^{r_c}dr \dfrac{r \cos^n(\theta(r))}{\sqrt{r^2-x^2}}
\end{equation}

Unfortunately, the integral in Eq.\ref{eq11} has no analytical solution in terms of simple functions. It is remarkable that the angular distribution keeps track of the power-law index $n$. In \cite{ref2} it was inferred, based on unpublished numerical
simulation that $\theta(r)$ as a function of $r$ is given by

\begin{equation}
\label{eq12} 
\theta(r)=\theta_0 \left(\dfrac{r}{r_c}\right)^2
\end{equation}

where $\theta_0\sim 43^\circ$. It can be seen that this dependence
satisfies the condition that the strand angle is small close to
the cable axis and large at the cable edge.

\section{Conclusions}
The analysis presented in this paper show that there are two limiting cases concerning the current transfer in cable in conduit conductors which influence the average procedure which must be performed when calculating different cable properties.
Let us follow the strand position as it meanders in the cable cross-section starting from a position at or very close to the $B_{min}$ position. The initial current in the strand is $I\leq I_c(B_{max})$. As the strand is moving in regions with higher field $B\geq B_{min}$ where $I\geq I_c(B)$ some current transfer ought to take place to neighboring strands. 
If the transversal inter-strand resistance is negligible small, the excess current, $\delta I=I-I_c(B)$ is easily transfered and the current in the strand follows nearly the variation of $I_c(B(x))$ in the cable cross-section. The consequence is that the cable cross-section is divided in vertical stripes where the field and the current in the strands are constant and the average is calculated with the distribution function from Eq.\ref{pxfinal}.

If the transversal resistance is very large, the surplus current can not be transfered and strands with an excess current will penetrate in the high field region of the cable. In this case the cable cross-section is divided in circular regions where the local magnetic field varies linearly between $B_0$ at the center and $B_{max}$ on the cable boundary. The average is calculated with a different distribution function Eq.\ref{pb2new}. 

The boundary value of the transversal resistivity, $\rho_t^*$ between the two regimes can be set by comparing the current-transfer length, Eq.\ref{eq2} with the length of the last twist pitch of the cable $p_S$

\begin{equation}
\label{comp}
L_{CT}=\left(\dfrac{0.1}{n}\right)^{\frac{1}{2}}\left(\frac{\rho_t^*}{\rho_c}\right)^{\frac{1}{2}}D_c=p_S
\end{equation}

Solving for $\rho_t^*$ one gets

\begin{equation}
\label{rho*}
\rho_t^*=10n\rho_c\left(\dfrac{p_S}{D_c}\right)^2
\end{equation}

As can be seen, if the inter-strand transversal resistivity $\rho_t\leq \rho_t^*$, the current redistribution is very effective. The limit value is proportional to $n$ the power -law index of the strands, increases with the square of the last twist pitch length and is inverse-proportional to the square of the cable diameter. Cables with large $n$, long last-stage twist-pitch length $p_S$ and small diameter $D_c$ have large $\rho_t^*$ values and better tolerance to higher transverse inter-strand contact resistance.

In both cases formulas for the calculation of the probability distribution function for the magnetic field are also presented.

A second issue treated in this paper is connected with the fact that the strands in a cable-in-conduit conductor do not cut the cable cross-section at right angles nor at any other constant angle. The cut angle is rather distributed, increasing monotonically from a small value for strands near the center of the cable to almost 90$^{\circ}$ at the cable border. The strand is therefore characterized not only by the position coordinates $x_i(z)$ and $y_i(z)$ in the cable cross section but also by the angle it has with a given transversal cut through the cable. The cable phase-space must be extended to the set $\left\{x_i(z), y_i(z), x'_i(z), y'_i(z)\right\}_{i\in \textbf{N}}$ with $x'_i=dx_i(z)/dz$ and $y'_i(z)=dy_i(z)/dz$.  The new distribution function, taking into account this effect is given in Eq.\ref{eq11}. The statistical mechanical approach sketched in this paper, based on the concept of viewing the strands as particles in movement could be very useful for the complete understanding of the complicated thermal and electromagnetic properties of cable-in-conduit conductors with twisted strands and multi-stage structure. We used here only one concept borrowed from statistical mechanics, the concept of ergodicity which allows one to replace the average over the length by an average over the cable cross-section.

\section*{Acknowledgment}
This paper is dedicated to the memory of Dr. G.~Pasztor for his continuing encouragement and support in the earlier 90's when I start working in this field and developed some of the concepts and points of view presented in this the work.

\end{document}